# Enhanced radiation tolerance of YSZ at high temperature against swift heavy ions: key role of interplay between material microstructure and irradiation temperature

Parswajit Kalita[a,*], Santanu Ghosh[a], Udai B. Singh[a,#], Pawan K. Kulriya[b], Vinita Grover[c], Rakesh Shukla[c], A.K. Tyagi[c], Gaël Sattonnay[d], Devesh K. Avasthi[e]

[a]*Department of Physics, Indian Institute of Technology Delhi, New Delhi – 110016, India.*

[b]*Materials Science Group, Inter University Accelerator Centre, New Delhi – 110067, India.*

[c]*Chemistry Division, Bhabha Atomic Research Centre, Mumbai – 400085, India.*

[d]*LAL, Université Paris-Sud, Bât. 200 F-91405, Orsay Cedex, France.*

[e]*Amity Institute of Nanotechnology, Amity University, Noida – 201313, India.*

**Abstract:**

Yttria stabilized Zirconia (YSZ) pellets with different crystallite sizes were irradiated with 80 MeV $Ag^{6+}$ ions at room temperature and 1000 K to understand the effect of crystallite size/material microstructure and irradiation temperature on the radiation tolerance against high electronic energy loss ($S_e$). X-ray diffraction and Raman spectroscopy measurements reveal that, irrespective of the irradiation temperature, the nano-crystalline samples suffered more damage as compared to the bulk-like sample. A reduction in the irradiation damage, i.e. improvement in the radiation tolerance, was observed for all the samples irradiated at 1000 K. The reduction in the damage, however, was remarkably higher for the two nano-crystalline samples compared to the bulk-like sample, and hence the difference in the damage between the bulk-like and nano-crystalline samples was also significantly lower at 1000 K than that at room temperature. The irradiation damage, against $S_e$, was thus found to be critically dependent on the interplay between the irradiation temperature and crystallite size. These results are explained with the help of detailed theoretical calculations/simulations based on the 'in-elastic thermal spike' model by taking into consideration the combined effect of crystallite size and environmental (irradiation) temperature on the electron-phonon coupling strength and lattice thermal conductivity (and hence on the resulting thermal spike). Our results are crucial from the fundamental perspective of comprehending the size and temperature dependent radiation damage against $S_e$; and also for a number of applications, in various radiation environments, where nano-materials are being envisioned for use.

*Correspondence email: phz148112@physics.iitd.ac.in

#Present address: *Department of Physics, Deen Dayal Upadhyaya Gorakhpur University, Gorakhpur – 273009, India.*

1.  **Introduction**

Materials are subjected to severe irradiation with energetic particles in a wide variety of environments e.g. in nuclear reactors, nuclear waste disposal systems, space vehicles etc. Such exposure of materials to irradiation is usually a detrimental process since it often results in defect creation leading to micro-structural changes (loss in crystallinity, swelling etc.) and eventually to the degradation of the materials' properties (embrittlement, cracking etc.) and thus a reduction of the operational life-time/efficiency of facilities. This becomes a crucial issue in the field of nuclear energy generation where materials are under constant bombardment by low energy neutrons and alpha recoil particles (typical energies of few hundred keV, lose energy in the target material by elastic collision with target atoms - referred to as nuclear energy loss ($S_n$)), high energy fission fragments (typical energy 70 – 100 MeV, lose energy in the target material by excitation and ionization of target atoms - referred to as electronic energy loss ($S_e$)) etc. It is therefore vital, especially with the rapidly increasing demand for 'cleaner' nuclear energy in the light of depleting fossil fuel reserves and/or climate change, to understand the radiation response of materials for the design of highly radiation tolerant materials for future efficient nuclear energy systems.

In the context of nuclear materials, cubic yttria stabilized zirconia (YSZ) is considered to be a very promising material for use as the host material matrix in inert matrix fuels (IMFs) to burn minor actinides and plutonium and/or to immobilize high level nuclear waste [1, 2]. The use of IMFs is especially important since a major concern with conventional nuclear energy generation using Uranium as the fuel is the safe management/disposal of radioactive Plutonium, minor actinides and other nuclear waste. Thus in view of the importance of cubic YSZ, a number of irradiation studies have been performed wherein its radiation tolerance has been investigated under both low energy ($S_n$ dominant) and high energy irradiations ($S_e$ dominant) and at irradiation temperatures ranging from room temperature to as high as about 1500 K [2-13]. Results of particular importance are that (i) no amorphization has ever been reported, (ii) against low energy irradiations, a multi-step damage evolution mechanism was observed with the damage being accelerated/higher at higher irradiation temperatures [2, 3, 5] i.e. no damage annealing was observed at high irradiation temperatures, and (iii) on the other hand, under high energy irradiations the damage evolution was found to follow a single-step [5] while a threshold $S_e \sim 20$ keV/nm was found to be required for the formation of stable clusters of displaced atoms or dislocation loops co-existing with singly ionized oxygen vacancies [9]; in addition, a decrease in the irradiation damage at higher irradiation temperatures (damage annealing) was observed [7, 12].

Nano-structured materials, meanwhile, have attracted wide interest [14-20] and are leading to revolutionary advances in the field of science and technology due to their exotic and size dependent properties. Of late, nano-materials have started garnering significant attention in the context of designing highly radiation tolerant materials for nuclear energy systems too - it has been speculated that nano-structured materials with their relatively large volume-fraction of grain boundaries (GBs) and 'short' diffusion distances will be more radiation tolerant as compared to their bulk/micro-crystalline counterparts since grain boundaries act as 'sinks' for defects; and therefore, in principle, GBs can act as 'sinks' for irradiation induced defects (such as interstitials and vacancies) too thereby lowering the accumulation of defects and hence enhancing the radiation tolerance. True to speculations, the enhanced radiation tolerance of nano-materials has been recently observed both experimentally and via computational approaches for various materials [6, 21-24] e.g. nano-crystalline YSZ was found to have lower concentration of defects compared to larger grained ones after irradiation [6], nano-crystalline $MgGa_2O_4$ spinel showed amorphization resistance up-to doses of 96 dpa which is significantly larger than the dose required to amorphize bulk $MgGa_2O_4$ [21] etc. It has indeed been shown by molecular dynamics simulations that interaction between GBs and defects strongly reduces the radiation damage – the mechanism being that interstitial defects are preferentially absorbed by the GBs which are then re-emitted from the GBs and annihilate vacancies via recombination (healing mechanism) [25]. A crucial point worth mentioning here is that in all these reports [6, 21, 22] the irradiations were performed with low energy ions i.e. where energy loss is dominated by $S_n$ and defects were produced by collision cascades - as the collision cascades interacted with GBs, interstitials were preferentially absorbed and then re-emitted leading to the healing mechanism (in nano-materials) and hence enhanced radiation tolerance. However, since the mechanism of $S_e$ and $S_n$ (and hence the resulting damage creation mechanisms) are fundamentally very different, an immediate logical question that hence arises is whether the radiation tolerance behavior of nano-crystalline and bulk/micro-crystalline materials are similar under high energy ($S_e$ dominant) irradiations as well. Apart from the grain size, the temperature of the surroundings (i.e. environmental/irradiation temperature) is another important factor that determines the overall radiation induced damage [2, 12, 26, 27] e.g. nano-crystalline $Lu_2Ti_2O_7$ and nano-crystalline tungsten were found to possess better resistance to radiation ($S_n$ dominated) damage at room temperature compared to their coarse-grained counterparts; at elevated temperatures, however, coarse-grained $Lu_2Ti_2O_7$ and tungsten exhibited better and similar radiation resistance respectively compared to their nano-crystalline counterparts [26, 27]. Thus, another query that emerges from these observations and also from the fact that material properties are both size and temperature dependent, is if the effect of irradiation temperature in determining the overall damages (created by $S_e$) the same for both nano-crystalline and bulk materials.

Despite the fact that a large number of publications have been devoted to the investigation of the radiation tolerance against $S_e$ dominated irradiations, the radiation tolerance of nano-materials (against $S_e$) has not yet been thoroughly investigated. To the best of our knowledge, only a couple of reports [28, 29] exist wherein the material was in the nano-crystalline form; moreover, detailed quantitative reasoning/explanation of the results [28, 29] were lacking. Furthermore, these reports [28, 29] (which do not include cubic YSZ) were concerned with the effects of room temperature irradiation only and hence the role of temperature on the radiation damage in nano-materials (against $S_e$) has not been explored at all. Therefore there is an acute lack of systematic and quantitative understanding of the physical processes that govern the structural response of nano-crystalline materials, and its difference with bulk (i.e. role of grain boundary density/grain size on radiation stability), against high $S_e$ at and above room temperature (for all materials in general and YSZ in particular). Thus, with the response of nano-crystalline materials to $S_n$ already quite well understood, understanding the behavior of these materials to high $S_e$ may be the key towards developing highly radiation tolerant materials for use in different radiation environments.

In the present study, nano-crystalline YSZ pellets were irradiated with 80 MeV $Ag^{6+}$ ions ($S_e \gg S_n$) at room temperature and 1000 K to investigate the effect of high $S_e$ and irradiation temperature on the nano-crystalline state. For a systematic understanding of the dependence of radiation tolerance on the crystallite/grain size (microstructure) and irradiation temperature against high energy irradiations, bulk-like YSZ pellets (with larger crystallites) were also irradiated under the same conditions. These ions were also chosen because they could effectively simulate the damage produced by fission fragments due to the similar $S_e$ in YSZ [12], while 1000 K is the typical nuclear reactor temperature. X-ray diffraction and Raman spectroscopy were used to investigate the effects of irradiation. Detailed thermal spike simulations/calculations were performed, by taking into account the combined effect of size and irradiation (environmental) temperature on the electron-phonon coupling factor and lattice thermal conductivity, to interpret the experimental observations. The major finding of this study is that a subtle interplay between the crystallite/grain size and irradiation temperature plays a decisive role in determining the final irradiation damage and hence the radiation stability of a material.

2. **Experimental**

10 mol% YSZ nano-powder was synthesized by gel combustion method using zirconium oxynitrate and yttria as reactants and glycine as the fuel [12]. The powder so obtained was divided into three parts and compacted into pellets of diameter ~ 8 mm. The pellets were then heated at 600ºC for 6 hours (henceforth referred to as S600), 800ºC for 4 hours (henceforth referred to as S800) and 1300ºC for 84 hours (henceforth referred to as S1300) with the aim of obtaining different microstructures and crystallite sizes. Densities of S600, S800 and S1300 were found to be 80%, 85% and 93% of the theoretical density respectively as measured by Archimedes' principle.

The sintered pellets were irradiated at room temperature (RT) and 1000 K (HT) with 80 MeV $Ag^{6+}$ ions using the 15UD Pelletron accelerator at IUAC, New Delhi. Irradiations were carried out at the fluence of $10^{14}$ ions/cm$^2$ for both temperatures with the particle flux and current maintained at 6.25x10$^9$ ions/cm$^2$/sec and 6 nA respectively. Details of the irradiation setup can be found elsewhere [30]. Electronic energy loss ($S_e$), nuclear energy loss ($S_n$) and the projected range of the incident ions were estimated to be ~19.9 kev/nm, 0.2 keV/nm and ~ 7.2 µm respectively by SRIM [31] simulation code.

X-ray diffraction (XRD) and Raman spectroscopy measurements, of pristine and irradiated pellets, were performed to investigate the structural changes/damage induced by irradiation. The XRD patterns were recorded with Cu$_{K\alpha}$ radiation in standard θ-2θ geometry using a Philips X'Pert Pro diffractometer. The incidence angle was fixed at 1º to ensure that the probed depth (~ 270 nm) remains well within the region and where $S_e \geq 19$ keV/nm. Raman spectra were recorded in a Renishaw InVia Raman microscope using a 405 nm laser as the excitation source. Pristine samples were also characterized by field emission scanning electron microscopy (FESEM) [MIRA, TESCAN] for information on the surface morphology, microstructure and particle size; an electron beam energy of 25 keV was used to record the SEM images.

3. Results

   3.1 Pristine Samples

XRD patterns of pristine S600, S800 & S1300 samples are shown in Figure 1. It is evident from these patterns that the same crystalline phase is obtained in S600, S800 and S1300; this phase was verified to be the cubic phase (cubic fluorite structure) according to PCPDF # 82-1246. The XRD patterns also show that the peaks become sharper and narrower upon increasing the sintering temperature which can be attributed to increase in crystallite size. The Williamson-Hall method was used to calculate the crystallite size from the XRD peak broadening and the average crystallite sizes were estimated to be ~ 26 nm, ~ 35 nm and ~ 80 nm for S600, S800 and S1300 respectively.

Figure 1 also shows representative SEM images of the non-irradiated pellets. It is clearly seen that S1300 is highly dense and has very well-defined particles of size ~ 4.5 ± 2.4 µm. This sample can thus be treated as bulk. In contrast, samples S800 and S600 are less dense and have much smaller particle sizes of ~ 38 ± 9 nm and ~ 57 ± 15 nm respectively. The apparent variation in the microstructure for the three samples is attributed to the different heat treatments that were provided.

   3.2 X-ray diffraction (irradiated samples)

The evolution of the XRD patterns for the samples irradiated at RT and HT are depicted in Figure 2. These XRD patterns reveal that, irrespective of the crystallite size/microstructure and irradiation temperature, the irradiation did not result in amorphization or any other major structural change. However, it is obvious from this figure that the XRD peak broadening has increased after irradiation along with reduction in the (integrated) peak intensities thus indicating that the long range crystallinity and periodic structure is being compromised i.e. irradiation has resulted in the deterioration of the crystallinity (damage). Careful inspection of the XRD patterns (Figure 2) however reveal that the damage is lower at HT (as compared to that at RT) for all the three sets of YSZ samples i.e. the effect of irradiation is less pronounced at HT. This is especially evident when the (111) diffraction maxima for pristine and RT, HT irradiated samples are compared – the irradiation induced broadening and reduction in intensity are smaller at HT indicating lesser damage. The reduction in the irradiation induced damage (at HT compared to RT), i.e. improvement in the radiation tolerance at HT, is in good agreement with our previous work [12] and is attributed to a dynamic mechanism in which the hindrance caused by the higher environmental (irradiation) temperature to the rapid thermal quenching of the localized hot molten zones (arising from spike in the lattice temperature upon irradiation) results in lower damage. Now, since XRD peak broadening is a measure of the degradation in crystallinity, the irradiation induced damage can be quantified by the relative/percentage change in the full width half maxima (FWHM). The irradiation damage was calculated using the following equation:

$$damage = \frac{FWHM_{(111)\_irradiated} - FWHM_{(111)\_pristine}}{FWHM_{(111)\_pristine}} \quad \ldots\ldots\ldots (1)$$

where $FWHM_{(111)\_pristine}$ & $FWHM_{(111)\_irradiated}$ are the FWHM of the (111) diffraction peak for the pristine and irradiated samples respectively. The FWHM of the (111) peak for the pristine and irradiated S600, S800 and S1300 samples and the corresponding damage are summarized in Table 1. Apart from the reduction in damage at HT (which was described earlier), a couple of other important observations are also made: (i) firstly, at RT, the two nano-crystalline samples (S600, S800) suffered significantly more damage than the bulk-like sample (S1300) – this result being distinctly different to previous reports [6] investigating the dependence of radiation stability on the crystallite size against low energy irradiations, and (ii) secondly, the damage at HT is again lower for S1300 compared to S600 and S800; however, surprisingly, the reduction in damage (at HT) is notably higher for the two nano-crystalline samples compared to the bulk-like sample. Thus, the XRD results show vital and surprising dependence of the radiation response behavior on the sample microstructure and irradiation temperature which will be addressed, in detail, in connection with the in-elastic thermal spike model/simulations discussed later.

|  | $FWHM_{(111)}$ (degrees) | Damage (XRD) | $FWHM_{F_{2g}}$ (cm$^{-1}$) | Damage (Raman) | Asymmetry of $F_{2g}$ band |
|---|---|---|---|---|---|
| S600 Pristine | 0.53 ± 0.02 | ----- | 100.8 ± 1 | ---- | 1.9 |
| S600 RT | 0.68 ± 0.02 | 29% | 129.0 ± 1 | 28% | 2.3 |
| S600 HT | 0.61 ± 0.02 | 15% | 118.1 ± 1 | 17.2% | 2.15 |
|  |  |  |  |  |  |
| S800 Pristine | 0.45 ± 0.02 | ----- | 97 ± 1 | ----- | 1.8 |
| S800 RT | 0.56 ± 0.02 | 24% | 118.7 ± 1 | 22.4% | 2.1 |
| S800 HT | 0.51 ± 0.02 | 13% | 108.4 ± 1 | 11.8% | 2 |
|  |  |  |  |  |  |
| S1300 Pristine | 0.30 ± 0.02 | ----- | 85.8 ± 1 | ----- | 1.75 |
| S1300 RT | 0.34 ± 0.02 | 13% | 96.7 ± 1 | 12.7% | 2 |
| S1300 HT | 0.33 ± 0.02 | 10% | 93.5 ± 1 | 9% | 1.9 |

**Table 1:** FWHM of (111) XRD peak, FWHM & asymmetry of $F_{2g}$ Raman band and the damage as calculated from XRD and Raman spectroscopy for all samples.

### 3.3 Raman spectroscopy (irradiated samples)

Raman spectroscopy is a powerful tool for investigating the local micro-structure due to the strong sensitivity to phonon characteristics of the crystalline material [29]. With the aim of investigating the changes brought upon at the microscopic level in the YSZ samples upon irradiation, Raman spectroscopy measurements were carried out. Figure 3 shows the Raman spectra of unirradiated and RT and HT irradiated S600, S800 and S1300 samples. A broad asymmetric band centered at around 605 cm$^{-1}$ is seen in all the three unirradiated spectra (~ 603 cm$^{-1}$ for S800 and ~ 606 cm$^{-1}$ for S600 & S1300) which matches with that reported in literature for cubic YSZ and is assigned to the Raman active $F_{2g}$ mode of Zr-O vibration with $O_h^5$ symmetry [32, 33]. The asymmetry of the $F_{2g}$ band can be attributed to the presence of disorder in the YSZ system induced by the doping of Yttrium ($Y^{3+}$) ions into the Zirconia structure [33].

The FWHM of the $F_{2g}$ band for pristine and irradiated S600, S800 and S1300 samples are listed in Table 1. Since the width of the Raman line is a measure of the crystallinity, it is apparent that irradiation (at RT & HT) has resulted in degradation of the crystallinity (peak broadening), i.e. damage, and that the outcome of irradiation is less prominent at HT. The damage was quantified using equation (1) with the FWHM of $F_{2g}$ band of pristine and irradiated sample (Table 1). Moreover, since line broadening on the low frequency side is typical for disordered systems, the ratio between the low and high frequency sides of the $F_{2g}$ line gives an estimate of the structural disorder [33]. Table 1 also shows the ratio of the widths of the low and high frequency sides at the half maxima position (asymmetry) of the $F_{2g}$ line, i.e. the structural disorder, for all the samples. In agreement with the XRD results, it can be again seen that: (i) the nano-crystalline samples (S600 & S800) are significantly more damaged (degradation in crystallinity and/or increase in structural disorder) at RT compared to the bulk-like (S1300) sample, and (ii) the reduction in damage, at HT, is appreciably larger for the nano-crystalline samples compared to the bulk-like sample. Additionally, it can also be seen in Figure 3 that, upon irradiation at RT, there is a red shift of the $F_{2g}$ band which increases from S800 to S600 - the $F_{2g}$ band for RT irradiated S800 and S600 are found at 588.2 cm$^{-1}$ and 585.8 cm$^{-1}$ respectively in contrast to the respective pristine samples for which the $F_{2g}$ band is at ~ 605 cm$^{-1}$. The red shift in a peak results from a decrease in the force constant. Since the RT irradiated S600 and S800 samples suffered significant irradiation damage, the tremendous red shift observed here can be explained as a manifestation of this damage i.e. the irradiation has resulted in the weakening of the Zr-O bond leading to a decreased force constant and hence a red shift of the $F_{2g}$ band. For the HT irradiated S800 and S600 samples, the $F_{2g}$ band is found at 593.7 cm$^{-1}$ and 591.1 cm$^{-1}$ respectively, again indicating reduced irradiation damage at HT. For S1300, the $F_{2g}$ peak is found at 603.5 cm$^{-1}$ and at 605 cm$^{-1}$ for the RT and HT irradiated samples i.e. in comparison to S600 & S800, the peak shift observed for the irradiated S1300 samples is negligible.

### 3.4 Thermal spike calculations/simulations

Computational simulations based on the 'inelastic' thermal spike model [34-36] were carried out in an attempt to understand the reasons/mechanism behind the dependence of the radiation tolerance on the microstructure and irradiation temperature. The 'inelastic' thermal spike model considers the target material to be composed of two separate sub-systems: namely, the electronic sub-system and the lattice sub-system. The energy of the incident ions is initially deposited in the electronic sub-system which is then shared among the electrons resulting in rise of the electronic temperature. Subsequently, the energy is transferred onto the lattice sub-system via electron-phonon (e-ph) coupling which results in a spike in the lattice temperature in a localized cylindrical region around the ion path. As a result, when the temperature surpasses the melting point of the material, a localized molten zone of few nanometers is formed which then gets rapidly quenched by thermal conduction. This transient local heating followed by the rapid quenching results in the formation of a damaged cylindrical region around the ion path, commonly referred to as the 'ion track'. The damaged cylindrical region/ion track maybe continuous or discontinuous or may even be a cluster of defects. This entire process, i.e. the rise and fall in the electronic and lattice temperature, is governed by two coupled differential equations in cylindrical geometry, given by

$$C_e(T_e)\frac{\partial T_e}{\partial t} = \frac{1}{r}\frac{\partial}{\partial r}\left[rK_e(T_e)\frac{\partial T_e}{\partial t}\right] - g(T_e - T_a) + A(r,t) \quad \ldots\ldots (2a)$$

$$\rho C_a(T_a)\frac{\partial T_a}{\partial t} = \frac{1}{r}\frac{\partial}{\partial r}\left[rK_a(T_a)\frac{\partial T_a}{\partial t}\right] + g(T_e - T_a) \quad \ldots\ldots (2b)$$

where, $C_e$, $C_a$, $K_e$ and $K_a$ are the specific heats and thermal conductivities of electronic and lattice subsystems respectively; $\rho$ is the material density; $T_e$ and $T_a$ are the electronic and lattice temperatures; $A(r,t)$ is the energy density supplied by the incident ions to the electronic subsystem; $g$ is the electron phonon coupling constant; $r$, $t$ are the radius of ion track and time respectively. The electron-phonon coupling constant ($g$) governs the efficiency of the energy transfer from the electrons to the lattice and hence determines the temperature spike in the lattice; while the heat/energy dissipation in the lattice, after the thermal spike, is governed by the lattice thermal conductivity ($K_a$). It is thus obvious that the rise, fall and duration of the lattice temperature and hence the irradiation induced damage crucially depends on the electron-phonon coupling factor and the lattice thermal conductivity. The electron-phonon coupling factor and the lattice thermal conductivity are expected to change with grain/crystallite size and/or environmental (irradiation) temperature, which can explain the observed dependence of the radiation response behavior of the YSZ samples on the size and temperature. It is therefore necessary to carefully consider the size and/or temperature dependence of '$g$' and '$K_a$' for the proper understanding of the complex radiation tolerance exhibited by the YSZ samples.

The electron-phonon coupling constant (which governs the efficiency of the energy transfer from the electrons to the lattice) is given by the relation [37]

$$g = \frac{9 n_e k_B^2 T_D^2 v_F}{16 \lambda T E_F} \quad \ldots\ldots\ldots (3)$$

In this relation, $n_e$ is the number density of electrons, $k_B$ is Boltzmann's constant, $T_D$ is the Debye temperature, $v_F$ is the Fermi velocity, $E_F$ is the Fermi energy, $T$ is the environmental temperature and $\lambda$ is the mean diffusion length of the excited electrons i.e. $\lambda$ is the electron-phonon mean free path. It is clear from eqn. (3) that $\lambda$ is the key parameter that determines the strength of the electron-lattice interaction and hence the temperature spike in the lattice. Now it is well known that grain boundaries act as a barrier for electron transport. Hence, the mean diffusion length of the electrons ($\lambda$) is strongly influenced by the grain/crystallite size, since as the size decreases and/or the amount of grain boundaries increases, the grain boundary scattering increases which consequently results in reduction of the electron-phonon mean free path. The relation between the mean diffusion length $\lambda$ and the grain/crystallite size ($d$) is given by [38, 39]

$$\lambda = \frac{\alpha d (1-R)}{R} \quad \ldots\ldots\ldots (4)$$

where $\alpha$ is the angle between the material plane and the velocity vector of the electrons and $R$ is the reflection coefficient of electrons striking the grain boundaries. Thus, due to the smaller grain/crystallite sizes, $\lambda$ will be smaller in the nano-crystalline systems as compared to bulk systems, which will result in stronger $g$ and consequently higher density of deposited energy into the lattice (in nano-crystalline systems). In the thermal spike simulations, eqn. (4) was used to calculate $\lambda$ for the different YSZ samples. As the values of $\alpha$ and $R$ are not known, the following assumptions were made [39]: (i) $\alpha$ and $R$ are constant for all the samples (constant for all the sizes), and (ii) the sample with the largest size (S1300) behaves like bulk single-crystalline. Therefore, for S1300, $\lambda$= 4.5 nm [28]. Using this value of $\lambda$ for S1300, the values of $\lambda$ are calculated, using eqn. (4), to be ~ 2 nm and ~ 1.5 nm for S800 (with d = 35 nm) and S600 (with d = 26 nm) respectively. For simplicity, $\lambda$ was assumed to be independent of the irradiation temperature. Moreover, as seen from eqn. (3), the electron-phonon coupling factor also varies with the irradiation (environmental) temperature, i.e. the combined effect of both size and temperature are taken into account in eqn. (3).

Meanwhile, as mentioned earlier, the lattice thermal conductivity is another vital parameter that influences the thermal spike since it governs the heat dissipation in the lattice. Again, since, grain boundaries act as an obstacle for heat transport by scattering and confining the motion of phonons, the lattice thermal conductivity is strongly dependent on the grain/crystallite size. In particular, as the size decreases and/or the density of grain boundaries increases, the scattering and/or confinement of the phonons becomes significant leading to a decrease in the thermal conductivity. As such, the dissipation of the heat after the ionizing thermal spike will be less effective in nano-crystalline systems, compared to bulk (microcrystalline) systems, resulting in longer duration of the thermal spike for nano-crystalline systems. The effective thermal conductivity for a material with grain/crystallite of size 'd' can be described by [40]

$$K = \frac{\dfrac{K_0}{1+\Lambda_0/d^{0.75}}}{1+\dfrac{R_k}{d}\left[\dfrac{K_0}{1+\dfrac{\Lambda_0}{d^{0.75}}}\right]} \qquad \ldots\ldots (5)$$

where $K_o$ is the single crystalline thermal conductivity, $\Lambda_o$ is the single crystal phonon mean free path and $R_k$ is the Kapitza thermal resistance. It should be noted here that the decrease of thermal conductivity with decreasing grain/crystallite size has been used, sometimes in conjunction with the size dependence of electron-phonon coupling factor, to explain the unusually large electronic energy loss mediated sputtering [37, 39, 41] and higher $S_e$ induced damage (loss/degradation in crystallinity) [28] observed in different nano-dimensional materials. However, the irradiations in these studies [28, 29, 37, 39, 41] were carried out at room temperature only, and were hence not concerned with the combined effect of irradiation temperature and size on the thermal conductivity (and 'g') and the following irradiation induced effects (sputtering, deterioration in crystallinity etc.). To account for the effect of the irradiation (environmental) temperature on the thermal conductivity, and hence on the thermal spike and the resulting irradiation damage, eqn. (5) was carefully scrutinized. The phonon mean free path term in eqn. (5) is in-fact temperature dependent and is given by [42]

$$\Lambda_0 = \frac{20T_m a}{\gamma^2 T} \qquad \ldots\ldots (6)$$

where $T_m$ is the melting temperature, $a$ is the lattice constant and $\gamma$ is the Gruneisen constant. Combining eqns. (5) & (6), we see that the temperature dependence of the thermal conductivity comes into the picture via the phonon mean free path. The size and temperature dependence of the lattice thermal conductivity are thus both simultaneously taken into consideration in Eqn. (5). For YSZ, $\Lambda_o$ = 25 nm at RT (T=300K) and 8 nm at HT (T=1000K) with $T_m$ = 2988 K, $a$ = 0.513 nm, $\gamma$ = 2. Equation (5) was then used to calculate the thermal conductivity of S600 (with d = 26 nm), S800 (with d = 35 nm) and S1300 (with d = 80 nm) which were used in the thermal spike simulations. The values of the other parameters required in the simulation were taken to be the same as that considered in our earlier work [12].

The thermal spike simulations/calculations were used to estimate the evolution of lattice temperature upon irradiation. The simulation results are shown in Figure 4 where the time dependence of the lattice temperature are plotted for two radial distances from the ion path – the center of the ion track or the ion path itself (0 nm) and the track radius. In each case the track radius was taken to be the maximum radial distance that could reach the melting temperature. The values of the maximum temperature reached, duration of the thermal spike and track radius are summarized in Table 2. It must be mentioned here that the thermal conductivity is influenced by the presence of porosity as well. The thermal conductivity values were thus corrected to account for the different amount of porosities present in our samples by using the following relation [43, 44]

$$K_{porous}/K_{dense} = 1 - (4P/3) \quad \ldots\ldots (7)$$

where porosity $P = 1 - (\rho/\rho_{theor})$ with $\rho$ and $\rho_{theor}$ being the experimentally estimated (by Archimedes' principle) and theoretical densities respectively, and $K_{dense}$ being the value of thermal conductivity in the absence of any porosity (which was calculated using eqns. (5) and (6)). The thermal spike simulations were also performed using the value of thermal conductivities obtained from eqn. (7) while keeping all other parameters same as the previous simulations, and the results are shown in Figure 5. It is apparent from this figure that regardless of the inclusion or exclusion of the porosity related correction to the thermal conductivity, the same trend in lattice temperature evolution is followed for all sizes and irradiation temperature with negligible differences in the values of the maximum thermal spike temperature and duration between the two cases. In other words, the inclusion of the porosity corrections only has an insignificant effect on the thermal spike and thus the role of porosity was not considered for the subsequent analysis. The simulation results (Figure 4, Table 2) reveal the following:

(i) It is generally considered that the material gets amorphized within the ion track. If we consider the same here, it is expected that the samples should be amorphized at ~ $10^{12}$ ions/cm$^2$ - 4.5x$10^{13}$ ions/cm$^2$ where complete overlap of the ion tracks take place corresponding to the estimated track radius of 3.8 nm – 0.6 nm respectively. However, as amorphization did not occur at even $10^{14}$ ions/cm$^2$, it is clear that the ion track is not amorphous and instead has defects. Similar existence of defects along the ion path has been previously reported for YSZ [10]. The formation of defects upon irradiation is also in good agreement with the results of Constantini et al. [9] who had reported a threshold value of $S_e$ ~ 20 kev/nm required for the formation of stable clusters of displaced atoms or dislocation loops in YSZ; in the present case, $S_e$ is very close to 20 keV/nm and hence it is reasonable to expect similar defects. The formation of defects/defect clusters, after irradiation under similar conditions, was indeed observed by transmission electron microscopy in our earlier work [12]. Sickafus et al. [7] had also reported similar findings for YSZ irradiated with $S_e$ ~ 19 keV/nm.

(ii) For the RT irradiations, the maximum thermal spike temperature reached, track radius and the duration of thermal spike increases from S1300 to S800 and then to S600. These values are significantly lower for S1300 as compared to those for S600 and S800.

(iii) In-case of the HT irradiations, all the three parameters i.e. duration of the thermal spike, track radius and the maximum thermal spike temperature reached, decrease as compared to the corresponding RT values for all the YSZ samples. These values are again the least for S1300 followed by S800 and then S600. However, the magnitudes of the decrease in the thermal spike duration and the maximum temperature, while going from RT to HT, are significantly larger for the two nano-crystalline samples (S600, S800) compared to the bulk-like sample (S1300) – the magnitude of the decrease being highest for S600.

|  | Max. Temp. (at 0 nm) | Track radius | Duration (within track radius) |
|---|---|---|---|
| S600 RT | 13850 K | 3.8 nm | ~ 60 ps |
| S600 HT | 5780 K | 3.2 nm | ~ 6.5ps |
|  |  |  |  |
| S800 RT | 11120 K | 3.6 nm | ~ 30 ps |
| S800 HT | 4575 K | 2.8 nm | ~ 2.4ps |
|  |  |  |  |
| S1300 RT | 5590 K | 2.4 nm | ~ 4ps |
| S1300 HT | 2988 K | 0.6 nm | ~ 1.4 ps |

**Table 2:** Maximum temperature obtained, track radius and duration of thermal spike (temperature at/above melting point) within the track radius for irradiated samples.

## 4. Discussion

The damage in the samples, as observed in the XRD and Raman measurements, is thus a manifestation of the irradiation induced defects/defect clusters. It is then interesting to investigate the origin of the variation in the damage depending on the grain/crystallite size and/or irradiation temperature. At RT, scattering and confinement of electrons and phonons is much more prominent in the nano-crystalline samples (S600, S800) compared to the bulk-like sample (S1300) because of the small size and relatively large density of grain boundaries. As a result, the electron phonon mean free path ($\lambda$) is lower and hence the electron-phonon coupling strength is stronger for the nano-crystalline samples compared to the bulk-like sample; while, the thermal conductivity is higher for the bulk-like sample compared to the nano-dimensional samples. Thus, for S600 and S800, as a result of the combination of these two factors, namely (i) strong capability of energy transfer from the electronic system to the lattice (enhanced 'g') and (ii) poor dissipation of the energy in the lattice (low thermal conductivity), the thermal spike induces much higher transient temperatures with longer durations as compared to S1300 (see Table 2). Consequently, as evident in the XRD and Raman results, significantly larger damage is produced in the nano-crystalline samples compared to the bulk-like sample after the ionizing thermal spike. This result is in contrast to that of Dey et al. [6] who had observed higher concentration of defects, i.e. higher damage, for large grained samples (220 nm) as compared to nano-grained ones (25 – 40 nm). Thus although there seems to be an ambiguity between our result and that of Dey et al. [6] (and also others [21, 22]), it must be noted here that the irradiations in these cases were carried out with low energy ions and hence the damage production was predominantly due to elastic nuclear energy loss ($S_n$). In our case the defect/damage production is predominantly due to inelastic electronic energy loss. Therefore, there is actually no ambiguity in the results; in-fact, they show that the radiation stability of nano-materials is critically dependent on the energy of irradiations i.e. energy loss/damage production mechanism. Similar to RT, at HT too, the electron phonon coupling factor and thermal conductivity are again stronger and lower respectively for the nano-crystalline samples thereby ultimately resulting in greater damage compared to the bulk-like sample. Thus at a particular temperature (RT or HT), the 'size effect' i.e. size dependent electron phonon coupling strength and thermal conductivity can be held responsible for the variation in the irradiation induced damage.

Now, the difference in the radiation damage for any particular size based on the irradiation temperature needs to be accounted for. In our earlier work [12] we had postulated that the resistance inflicted by the high environmental (irradiation) temperature on the thermal spike is responsible for the lower damage produced at high irradiation temperature. However, for a quantitative understanding, the effect of temperature on the electron phonon coupling and lattice thermal conductivity must be considered. Firstly, it is clear from eqn. (3) that $g$ is lower at HT (T=1000K) compared to that at RT (T=300K). As such, the energy transferred to the lattice from the electrons (upon irradiation) will be lower at HT thereby resulting in lower transient lattice temperatures at HT. Secondly, it is logical to assume that reduction in thermal conductivity due to scattering of phonons by grain boundaries is significant only when grain/crystallite size ≤ phonon mean free path ($\Lambda_o$). Since, $\Lambda_o^{1000K} < \Lambda_o^{300K}$ (eqn. (6)), the scattering of phonons by the grain boundaries is less effective at 1000 K compared to 300 K thereby leading to better heat dissipation at 1000 K and hence decreased duration of thermal spike. Thus in addition to lower transient lattice temperatures, the thermal spike duration is also shorter at HT (see Table 2) leading to lesser damage.

A final discussion is devoted to the significant reduction in the irradiation damage (significantly enhanced radiation tolerance) at 1000 K for the nano-crystalline samples versus the slight damage reduction (slight enhancement in radiation tolerance) observed for the bulk-like sample. For any particular size, the electron phonon coupling strength is always lower at 1000 K than that at 300 K and hence lesser energy is transferred to the lattice at 1000 K.

For the bulk-like sample (d = 80 nm), the phonon mean free path at both 300 K ($\Lambda_o$= 25 nm) and 1000 K ($\Lambda_o$= 8 nm) is much less than the crystallite size. Since the scattering of phonons from grain boundaries is effective only when $\Lambda_o \sim d$, very little scattering can thus be expected at both temperatures (since d >> $\Lambda_o$ at both 300 K and 1000 K). Hence the heat dissipation is efficient and also effectively remains the same for both 300 K and 1000 K. In other words, the change (improvement) in the heat dissipation, at 1000 K, is not significant. Therefore, when going from 300 K to 1000 K, only the change (decrease) in the electron-phonon coupling strength ($g$) is effectively responsible for change in the thermal spike and hence the change (decrease) in the damage, with the change in the value of $\Lambda_o$ (improvement in the heat dissipation) playing a trivial role. This can be clearly seen from the thermal spike simulation results where the duration of the thermal spike only changes slightly from ~ 4ps at 300 K to ~ 1.4 ps at 1000 K (Table 2). As such, the irradiation damage at 1000 K is slightly lower than the damage at 300 K i.e. there is only a slight enhancement in the radiation tolerance.

On the other hand, in case of the nano-crystalline samples (d = 26 – 35 nm), the phonon mean free path at 300 K ($\Lambda_o$= 25 nm) is equivalent to the crystallite size; while d >>$\Lambda_o$ at 1000 K ($\Lambda_o$= 8 nm). Therefore, the scattering of phonons by the grain boundaries is efficient at RT leading to poor heat dissipation. In contrast, at 1000 K, the scattering is negligible and thus the heat dissipation is significantly better. This is in addition to the decrease of electron phonon coupling strength at 1000 K. Therefore as opposed to the bulk-like sample, both the change (reduction) in the value of $\Lambda_o$ (improvement in the heat dissipation) and change (decrease) in the electron phonon coupling strength play a significant role in changing the thermal spike, while going from 300 K to 1000 K, for the nano-crystalline samples. This ultimately results in much lesser damage (at 1000 K) i.e. results in significantly higher reduction of the irradiation induced damage (significant enhancement in radiation tolerance) for the nano-crystalline samples. This is again evident from the thermal spike simulations where both the maximum lattice temperature reached and thermal spike duration decreases considerably (Table 2).

A subtle interplay between the crystallite size (microstructure) and the irradiation (environmental) temperature therefore plays a key role in determining the eventual irradiation damage; and hence, the enhancement in the radiation tolerance at high temperature is also determined by the interplay between the material microstructure and irradiation temperature. A schematic of the damage production mechanism for nano-crystalline and 'bulk' samples, at 300 K and 1000 K, is shown in Figure 6.

Finally, since the electron phonon coupling strength is always higher for nano-crystalline samples, and the efficiency of heat dissipation in nano-crystalline samples is strongly temperature dependent (being efficient only at elevated temperatures) in contrast to bulk samples for which the heat dissipation is consistently efficient and is effectively independent of the temperature, the explanation predicts that at any particular irradiation temperature, bulk (micro-crystalline) samples will always be less damaged than nano-crystalline samples with the difference in the damage depending on the irradiation temperature. Irradiations over a wide range of sizes and temperatures are required to test this hypothesis.

## 5. Conclusions:

To summarize, YSZ pellets with different crystallite sizes (26 nm – 80 nm) were irradiated with 80 MeV $Ag^{6+}$ ions ($S_e \gg S_n$) at room temperature and 1000 K for a comprehensive understanding of the role of crystallite size/microstructure and irradiation temperature on the radiation tolerance against high $S_e$. No amorphization was observed irrespective of the crystallite size and irradiation temperature. At room temperature, the two nano-crystalline samples (26 nm, 35 nm) suffered significantly more damage as compared to the bulk-like sample (80 nm). This is attributed to the combination of stronger electron phonon coupling and lower thermal conductivity, both of which are a consequence of the smaller crystallite/grain size, for the two nano-crystalline samples (as compared to the bulk-like one) which results in much higher transient lattice temperatures with longer durations and hence much higher damage. A reduction in the irradiation damage i.e. improvement in the radiation tolerance, was observed for all the samples irradiated at 1000 K. Similar to that at RT, the nano-crystalline samples were again more damaged than the bulk-like one. The reduction in the damage, however, was remarkably higher for the two nano-crystalline samples compared to the bulk-like sample, and hence the difference in the damage between the bulk-like and nano-crystalline samples was also significantly lower at HT (than that at RT). Given the 'small' size of the nano-crystalline samples and temperature dependence of the phonon mean-free path, the heat dissipation in the lattice is efficient only at 1000 K. In contrast, for the bulk-like sample, the lattice heat dissipation is always efficient and effectively temperature independent because of its 'large' size. Therefore, as opposed to the bulk-like sample for which the decrease in the electron-phonon coupling, at 1000 K, is the only significant factor responsible for the decrease in the irradiation damage, both the decrease in electron-phonon coupling and improvement in heat dissipation, at 1000 K, are significant for the nano-crystalline samples which results in much higher damage reduction. The irradiation damage, against $S_e$, is thus critically dependent on a subtle interplay between the irradiation temperature and grain size. Finally, besides the fact that our results are of fundamental importance from the perspective of advancing the understanding of ($S_e$ dominated) ion – matter interaction, they may also have important implications for the use of nano-crystalline materials in different radiation environments (particularly the nuclear industry) where $S_e$ dominated radiations and different temperatures are encountered.

**Acknowledgements:**

We sincerely acknowledge Prof. Pankaj Srivastava, IIT Delhi for helpful discussions. We express gratitude towards Dr. Saif A. Khan and all the members of Pelletron group, IUAC New Delhi for their help during SEM measurements and irradiation respectively. Parswajit Kalita is thankful to MHRD, India (IIT Delhi) and IFCPAR/CEFIPRA for financial assistantship. XRD (Department of Physics) and Raman spectroscopy (FIST UFO scheme) facilities of IIT Delhi are acknowledged. Devesh K. Avasthi is thankful to DST, India for funding the SEM facility under Nano Mission project.

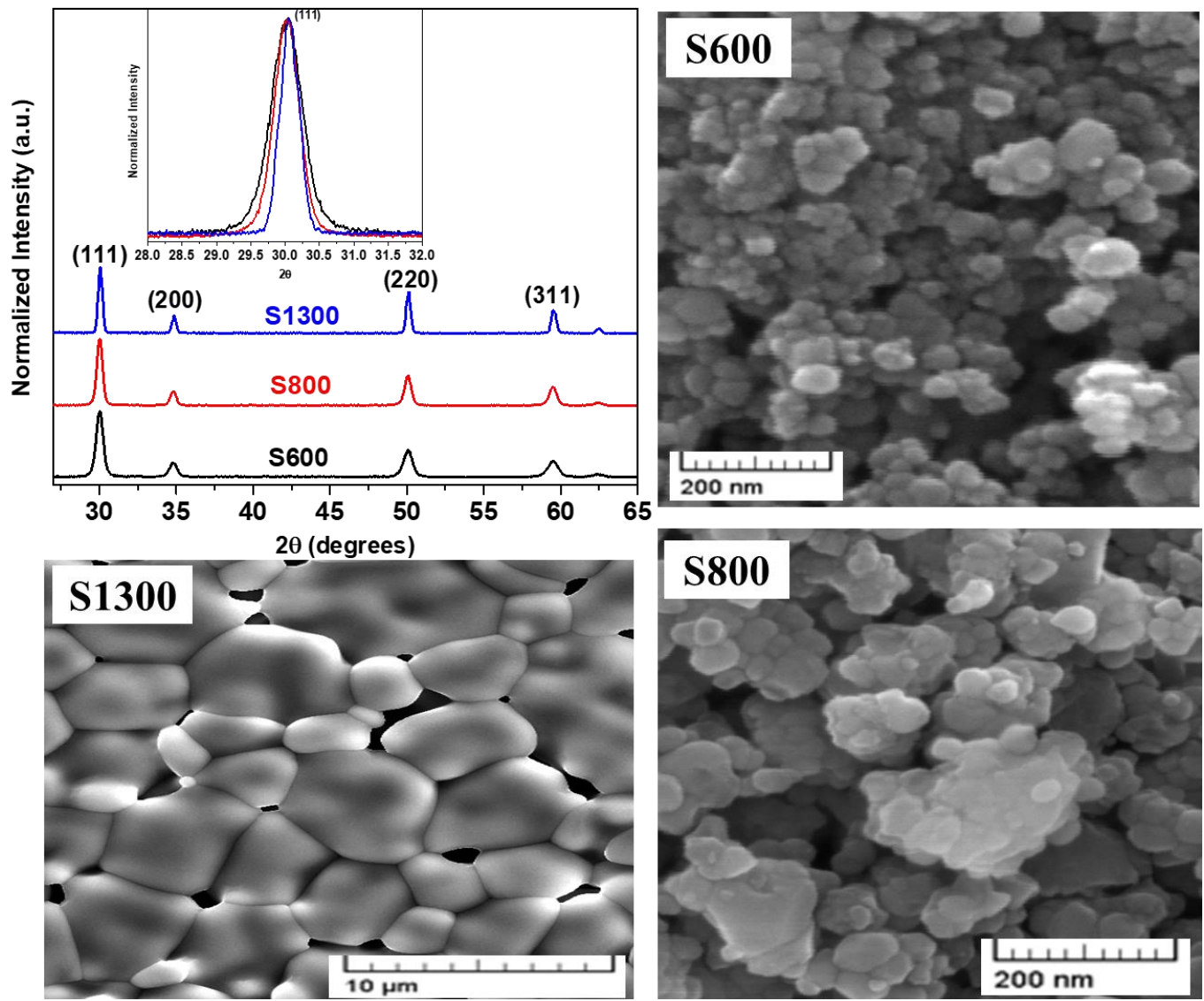

**Figure 1:** *(in clockwise direction)* GIXRD patterns of pristine S600, S800 & S1300, magnified view of (111) peak for all the samples is shown in the inset; SEM images of pristine S600, S800 & S1300.

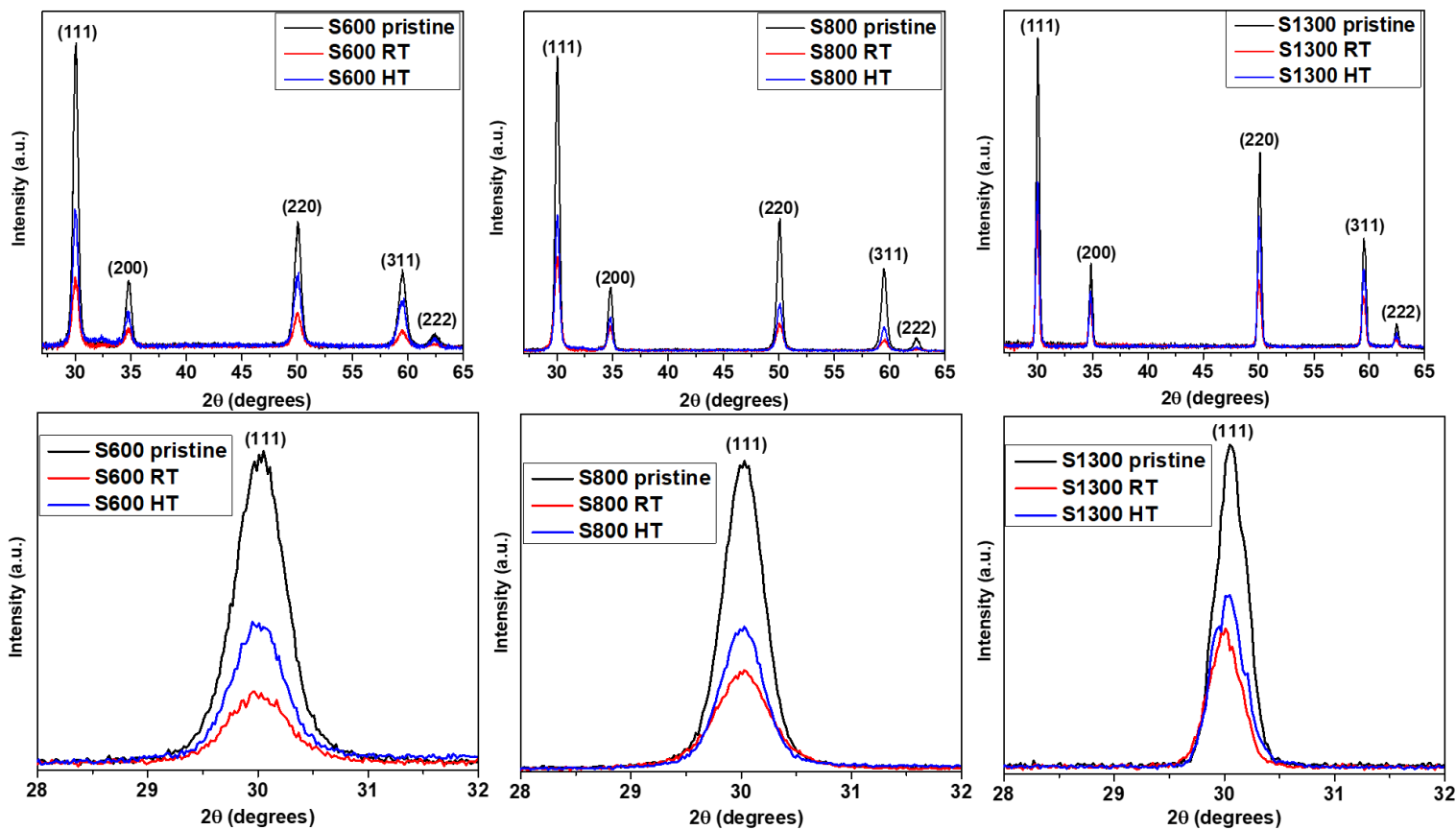

**Figure 2:** GIXRD patterns of pristine and irradiated S600, S800 & S1300. Magnified (111) peak for all the samples are also shown. For each particular size, the integrated intensity of the (111) diffraction maxima, of the pellets used, were equal (within statistical fluctuations) before irradiation.

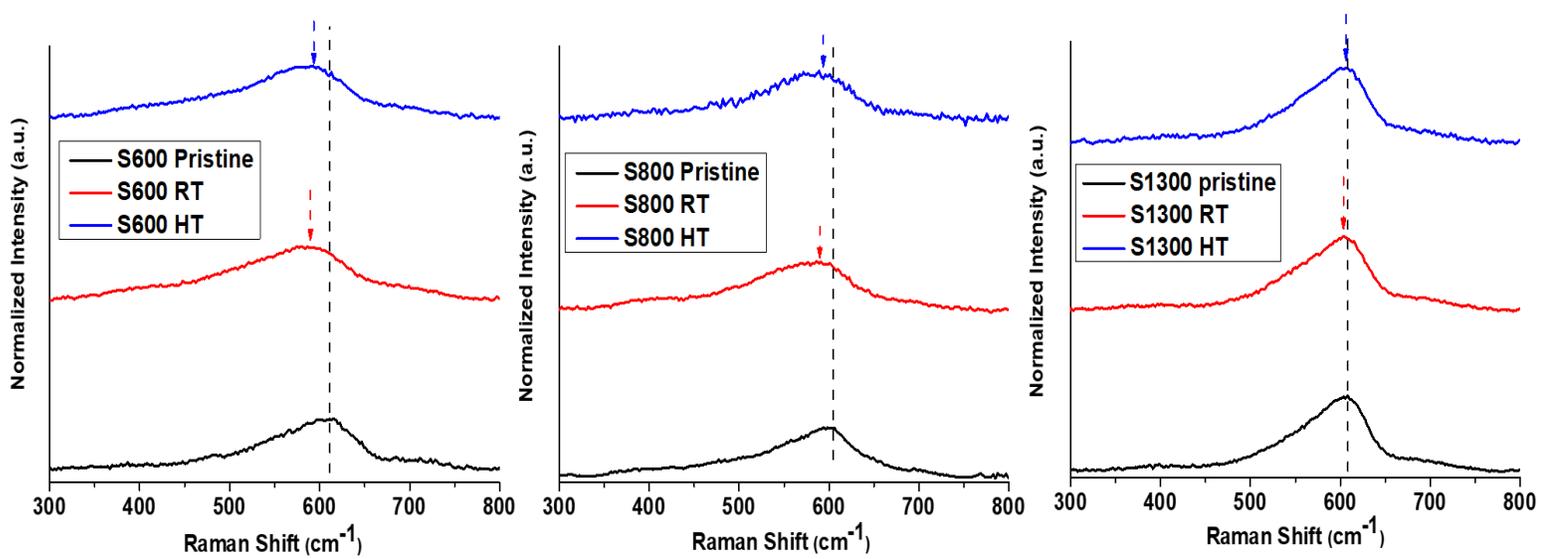

**Figure 3:** Raman spectra of pristine and irradiated S600, S800 & S1300 samples showing the $F_{2g}$ band. Arrows indicate the peak position, the dashed line is given for the clear visualization of the peak shift after irradiation.

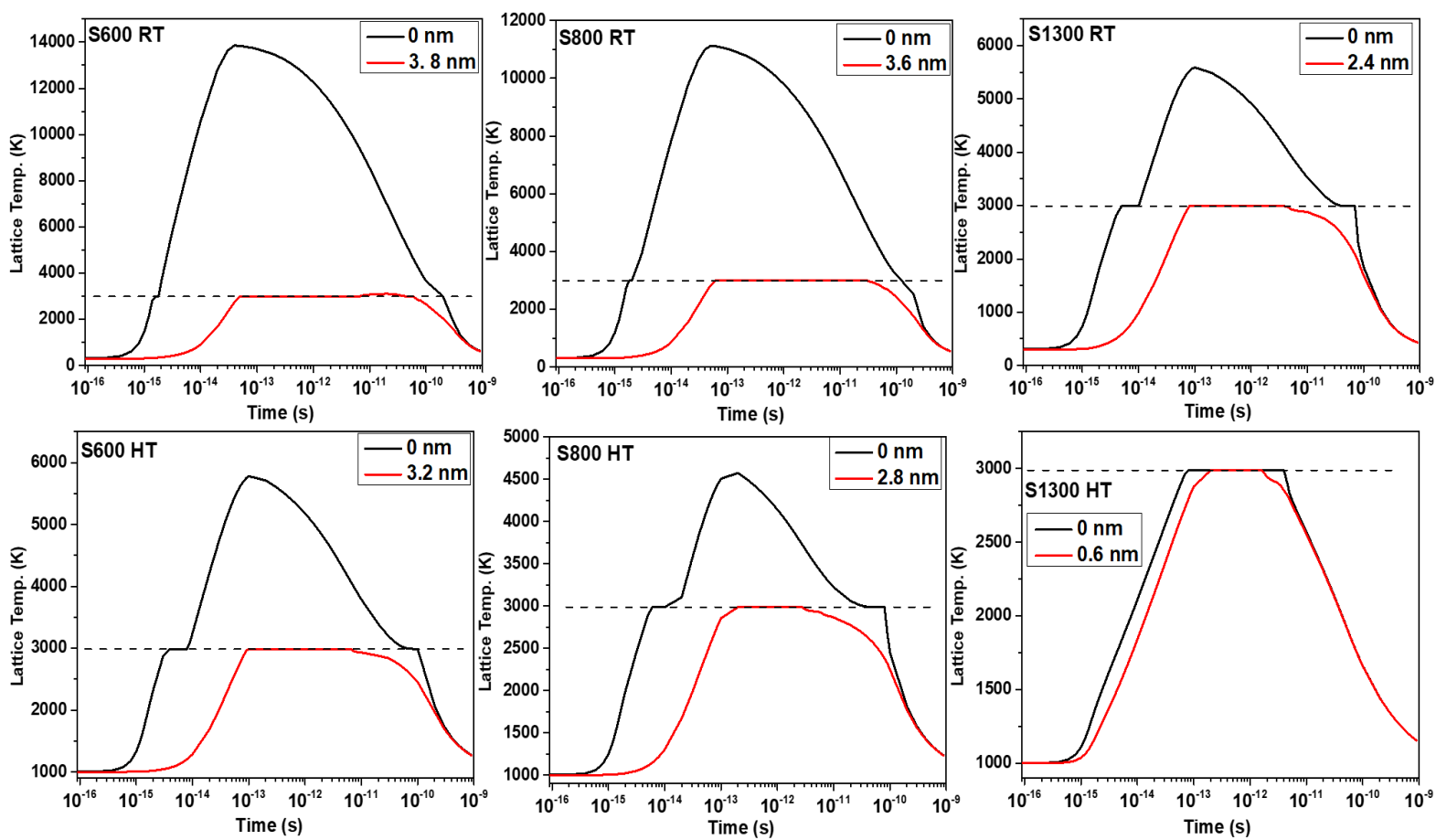

**Figure 4:** Variation in lattice temperature with time at the center of ion track (0 nm) and at respective track radius. The dashed horizontal line shows the melting temperature (2988 K).

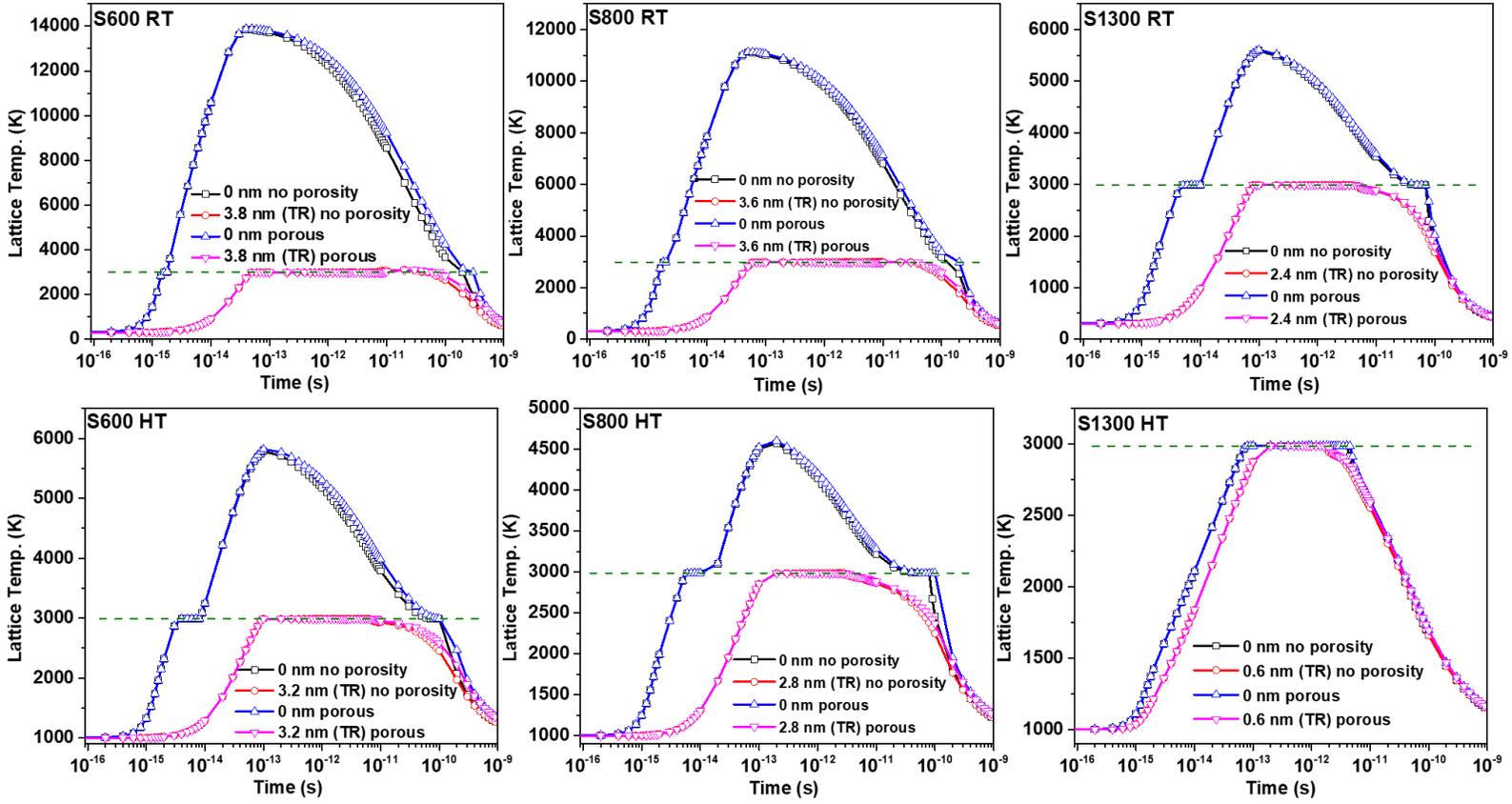

**Figure 5:** Variation in lattice temperature with time at the center of ion track (0 nm) and at respective track radius (TR), both with and without porosity corrections. The dashed horizontal line shows the melting temperature (2988 K).

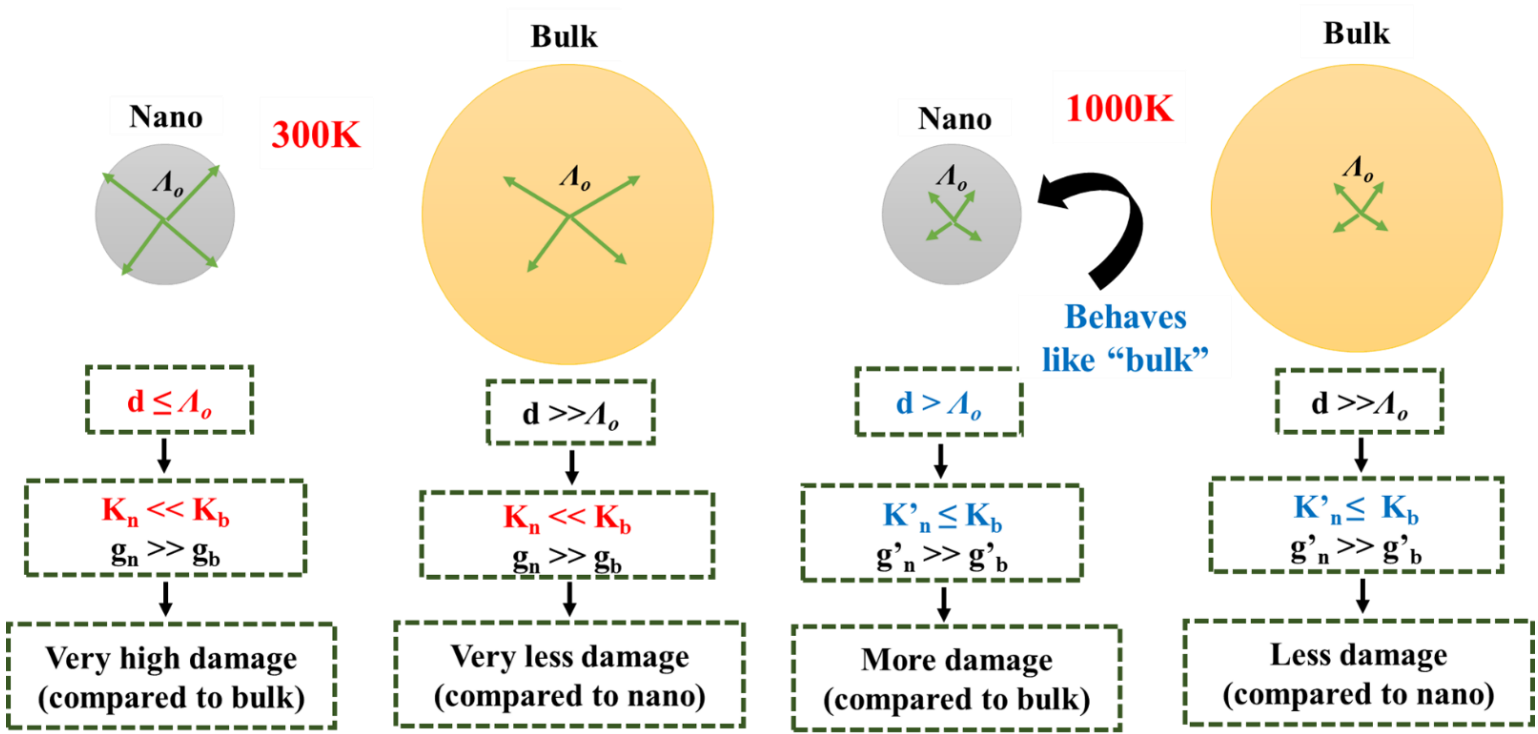

**Figure 6:** Schematic of the proposed dependence of radiation damage on crystallite/grain size and irradiation temperature. $K_n$, $K_b$, $g_n$, $g_b$ are the lattice thermal conductivity of nano-crystalline, lattice thermal conductivity of bulk, electron-phonon coupling factor for nano-crystalline and electron-phonon coupling factor for bulk samples respectively at 300 K; $K'_n$, $K'_b$, $g'_n$, $g'_b$ are the lattice thermal conductivity of nano-crystalline, lattice thermal conductivity of bulk, electron-phonon coupling factor for nano-crystalline and electron-phonon coupling factor for bulk samples respectively at 1000 K.